# Low-Temperature Characteristics of an AlN/Diamond Surface Acoustic Wave Resonator

Moyuki Yamamoto,[1] Hodaka Kurokawa,[2] Satoshi Fujii,[3,4] Toshiharu Makino,[2,4] Hiromitsu Kato[2,4], and Hideo Kosaka[1,2]

[1]*Graduate School of Engineering Science, Yokohama National University, 79-5 Tokiwadai, Hodogaya, Yokohama 240-8501, Japan*

[2]*Quantum Information Research Center, Institute of Advanced Sciences, Yokohama National University, 79-5 Tokiwadai, Hodogaya, Yokohama 240-8501, Japan*

[3]*Research Center for Electronic and Optical Materials, National Institute for Materials Science, 1-1 Namiki, Tsukuba, Ibaraki, 305-0044, Japan*

[4]*Advanced Power Electronics Research Center, National Institute of Advanced Industrial Science and Technology, 1-1-1 Umezono, Tsukuba, Ibaraki, 305-8568, Japan*

Phonons confined in mechanical resonators can be coupled to a variety of quantum systems and are expected to be applied to hybrid quantum systems. Diamond surface acoustic wave (SAW) devices are capable of high efficiency in phonon interaction with color centers in diamond. The temperature dependence of the quality factor is crucial for inferring the governing mechanism of coupling efficiency between phonons and color centers in diamond. In this paper, we report on the temperature dependence of the quality factor of an AlN/diamond SAW device from room temperature to 5 K. The temperature dependence of the quality factor and resonant frequency suggests that the mechanism of SAW dissipation in the AlN/diamond SAW resonator at 5 GHz is the phonon-phonon scattering in the Akheiser region, and that further cooling can be expected to improve the quality factor. This result provides a crucial guideline for the future design of AlN/diamond SAW devices.

I.    **INTRODUCTION**

Phonons confined within mechanical resonators can be coupled to various quantum systems such as superconducting qubits[1], photons[2], and electrons in solids[3–5], making them promising for applications to hybrid quantum systems. Among them, optically active color centers in diamond have been proposed to operate as transducers between microwaves and light when coupled with microwave phonons[6,7]. Also, the manipulation of color centers by phonons is itself an interesting problem. Color centers with spin-orbit interactions in the ground state, like SiV, exhibit the ability to manipulate spin states at low power by leveraging orbital degrees of freedom that have a strong coupling to phonons. Demonstrations of electron and nuclear spin manipulations utilizing phonons have indeed been achieved[4,8].

For phonon-based color center manipulation, cantilever structures[9–11], bulk acoustic wave devices[12], and diamond surface acoustic wave (SAW) devices[4,13] are used to generate phonons. SAW resonators can handle several GHz to a dozen GHz, exhibit relatively high quality factors, and can reduce the mode volume[3,5], allowing efficient interaction between phonons and color centers. The higher the quality factor of the resonator, the more efficient the interaction between phonons and the color center, so the quality factor must be as high as possible to manipulate the color center at low power and to achieve a strongly coupled state between phonons and the color center. To assess the governing mechanism behind the upper limit of the quality factor in diamond SAW devices, it is useful to measure the temperature dependence of the quality factor. However, there are no systematic reports on the temperature dependence of the quality factor of widely used AlN/single-crystal diamond SAWs from low temperatures (below 20 K) to room temperature.

In this study, we measured the temperature dependence of the quality factor of a 1-port AlN/diamond SAW device from room temperature to 5 K. The temperature dependence of the quality factor and resonant frequency suggests that the AlN/diamond SAW device is in the Akheiser region, and that the phonon-phonon scattering has a dominant role in determining the quality factor from room temperature to around 5 K. The quality factor was not saturated at 5 K, and further improvement in it can be expected by cooling the device below 5 K. These results will be of great help in guiding the design and operation of AlN/diamond SAW devices in the future.

## II. METHODS

The design and fabrication method of the 1-port SAW resonator used in the experiment was the same as that proposed by Fujii et al[14]. Polycrystalline AlN thin film with *c*-axis orientation was deposited by RF magnetron sputtering on type Ib single-crystal diamond with (100) orientations, which was produced by a high-temperature, high-pressure method (Sumitomo Electric). Electrode patterns were formed by electron beam lithography. The design of resonator was oriented in the diamond crystal direction <010> and consisted of 90 pairs of IDTs with $\lambda$ = 2.0 μm wavelength and 90 finger reflectors on each side of the IDTs. The AlN film thickness was 0.7 μm (KH = 0.35), and the Al electrodes thickness were 90 nm. Based on theoretical calculations, Sezawa waves were excited, with a phase velocity of 10.5 km/s (center frequency of 5.25 GHz). The electromechanical coupling constant $k^2$ was designed to be 1.2%.

Figure 1(b) is a schematic of the experimental setup: the SAW resonator was attached to a printed circuit board with varnish (GE7031 varnish). The coplanar waveguide and SAW resonator electrodes on the printed circuit board were wirebonded with aluminum wires. The signal lines on the printed circuit board were routed through SMP connectors. Using a vector network analyzer (Keysight, P9373A), the reflection coefficients, $S_{11}$, were measured while the sample was mounted on a cryostat (Montana Instruments, Cryostation® s100). To prevent nitrogen from solidifying and adhering to the sample at low temperatures, a turbo pump was used to draw a vacuum to less than 0.1 mPa before cooling. During cooling, the pump was stopped and disconnected from the cryostat. Measurements at low temperatures were made after waiting at least 10 minutes after the temperature sensor reached the target temperature.

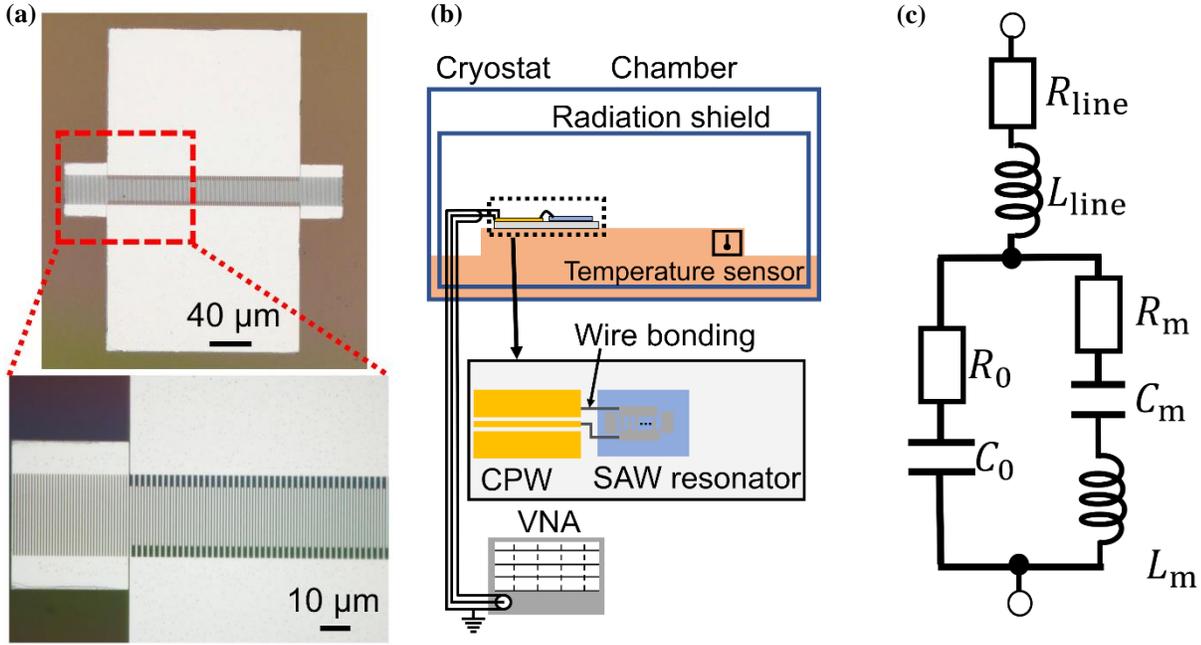

Fig. 1. (a) Image of SAW resonator used in this experiment. The IDTs and reflectors made of Al were fabricated on the AlN/diamond heterostructure. (b) Schematic of the S-parameter measurement setup: the SAW resonator is attached to a printed circuit board, and the coplanar waveguide (CPW) and electrodes on the board are coupled by wire bonding. The sample is cooled to 5 K by a cryogen-free cryostat. (c) mBVD-model with wire and electrode impedances connected in series.

Before cooling the sample, we used homemade calibration kits (OPEN, SHORT, LOAD) fabricated on the printed circuit boards. Since our calibration kits cannot be calibrated for the circuit after the coplanar waveguide, the effects of the Al wires and electrodes in addition to the SAW resonator remained in $S_{11}$. The equivalent circuits of the Al wire, electrode, and SAW resonator are modeled using the modified Butterworth-Van Dyke (mBVD) model with the composite resistance of the electrode and wire $R_{\text{line}}$ and inductance $L_{\text{line}}$. In Fig. 1 (c), $C_0$ is the composite capacitance of the electrode and wire; $R_0$ is the radiation loss and the loss from the Al electrode and wire before the ground; and $R_{\text{m}}$, $L_{\text{m}}$, and $C_{\text{m}}$ are respectively the motional resistance, motional inductance, and motional capacitance of the SAW resonator. The impedance, $Z_{\text{DUT}}$, of this equivalent circuit can be expressed by the following equation using the characteristic impedance, $Z_0$, of the measurement system and $S_{11}$.

$$Z_{\text{DUT}} = Z_0 \frac{S_{11}+1}{S_{11}-1} . \tag{1}$$

$L_{\text{line}}$, $C_0$, and $R_0$ were determined from the characteristics over a wide frequency range other than the resonant frequency of the SAW. $R_{\text{line}}$ is the resistance considering the skin effect of the wire and electrode, and it was set to 1 Ω as a parameter. The remaining parameters, $R_{\text{m}}$, $L_{\text{m}}$, and $C_{\text{m}}$, were determined by fitting the admittance near the resonant frequency (5.2-5.3 GHz).

Though the absolute values of $R_m$, $L_m$, and $C_m$ were affected by the arbitrariness of the determination of $R_{line}$, their temperature dependence was not significantly affected in the range of 0.5-1.5 Ω, so it was not considered to affect the later discussion.

To eliminate the effect of impedance changes associated with decreasing temperature, the following procedure was used to remove the effect of temperature changes on $S_{11}$. The measured $S_{11}$, $S_{11}^{meas}$, can be expressed as

$$S_{11}^{meas}(T) = S_{11}^{DUT}(T) T^{cable}(T), \tag{2a}$$

$$\log|S_{11}^{meas}(T)| - \log|S_{11}^{DUT}(T)| = \log|T^{cable}(T)|, \tag{2b}$$

$$\theta^{meas}(T) - \theta^{DUT}(T) = \theta^{cable}(T), \tag{2c}$$

where $S_{11}^{DUT}(T)$ is the $S_{11}$ of the whole device expressed using the mBVD model; $T^{cable}(T)$ is the temperature-dependent transmittance of the cable; and $\theta^{meas}(T)$, $\theta^{DUT}(T)$, and $\theta^{cable}(T)$ are the phases of $S_{11}^{meas}(T)$, $S_{11}^{DUT}(T)$, and $T^{cable}(T)$, respectively. Since $T^{cable}(T)$ is 1 at room temperature due to the calibration, $S_{11}^{meas}(T)$ and $S_{11}^{DUT}(T)$ have the same value at room temperature. Therefore, $\log|S_{11}^{meas}(T^{RT})| = \log|S_{11}^{DUT}(T^{RT})|$, and $\theta^{meas}(T^{RT}) = \theta^{DUT}(T^{RT})$. At low temperatures, $S_{11}$ away from the resonant frequency is governed by the temperature variations of $R_{line}$, $L_{line}$, $R_0$, $C_0$, and the cable. However, since these are independent of the resonance characteristics, the effect of temperature variation can be attributed entirely to the cable. Under these assumptions, the following equation holds,

$$\log|S_{11}^{meas}(T)| - \log|S_{11}^{meas}(T^{RT})| = \log|T^{cable}(T)|, \tag{3a}$$

$$\theta^{meas}(T) - \theta^{meas}(T^{RT}) = \theta^{cable}(T). \tag{3b}$$

Using equation (3a) and (3b) at frequencies far from the SAW resonance frequency, we can estimate the temperature-dependent change of the cable impedance. Using the obtained $\log|T^{cable}(T)|$ and $\theta^{cable}(T)$, we can calibrate out the temperature-dependent shift in $S_{11}$ in the whole measurement frequency range assuming that the frequency dependences of $\log|T^{cable}(T)|$ and $\theta^{cable}(T)$ are weak. Impedance or admittance was calculated after the temperature was calibrated. Because of the assumption used here, we treat $R_{line}$, $L_{line}$, $C_0$ as constants in the following.

## III. RESULTS AND DISCUSSION

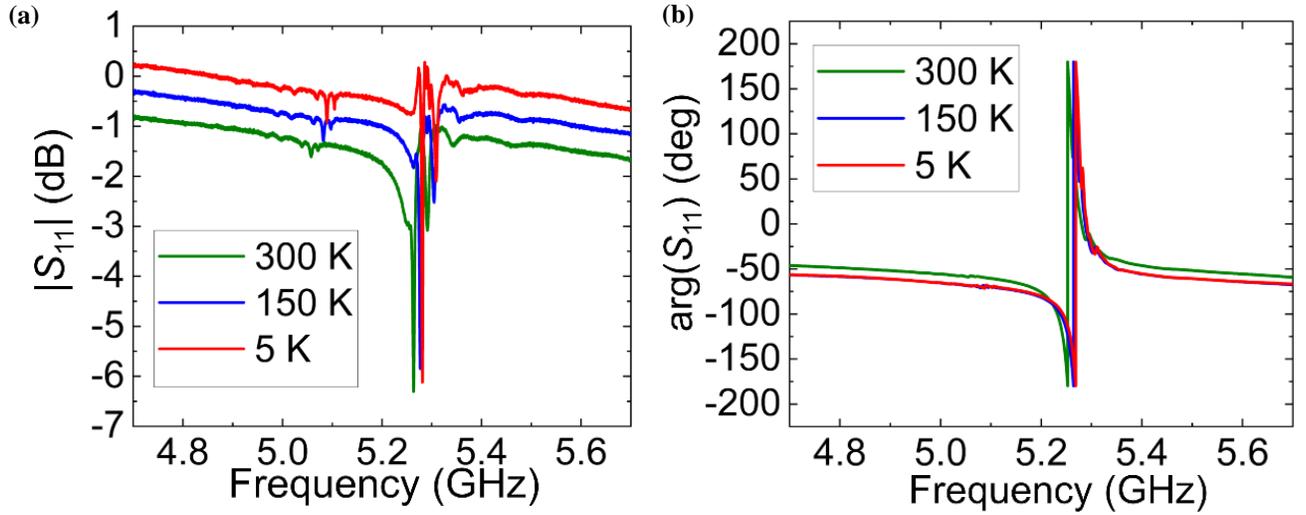

Fig. 2. Before the temperature correction of (a) magnitude and (b) phase of $S_{11}$. The wiring loss decreases with decreasing temperature, and the phase at the region away from the resonant frequency $|S_{11}|$ is decreasing.

Figures 2(a) and 2(b) show $S_{11}^{\text{meas}}$ and $\theta^{\text{meas}}$, respectively, before the temperature correction at each temperature. The temperature-dependent shift of $S_{11}^{\text{meas}}$ was uniform in the measurement frequency range. Therefore, our assumption described in the previous section for the calibration holds. After the correction of the temperature-dependent shift, $S_{11}^{\text{meas}}$ was converted to admittance and fitted using the mBVD model described in the previous section. Figure 3 shows the real and imaginary parts of the admittance at 300 K and 5 K. The linewidth of admittance was narrower when cooled to 5 K compared to that at room temperature. Table 1 summarizes the parameters obtained from the fitting. Figure 4(a) shows the mechanical quality factor, $Q = 2\pi f_r L_m / R_m$, at various temperatures. In the measurement range (5-300 K), the quality factor increased with decreasing temperature, indicating the decrease in dissipation. Figure 4(b) shows the temperature dependence of phase velocity, $v(T) = f_r/\lambda$. At $T > 50$ K, $v(T)$ increased with decreasing temperature, while at $T \leq 50$ K, it converged to around 10.56 km/s. $v(T)$ can be expressed as $v(T) = \sqrt{E(T)/\rho(T)}$, where $E(T)$ is the Young's modulus and $\rho(T)$ is the mass density. As the temperature of the SAW resonator is lowered, $E(T)$ and $\rho(T)$ are both expected to increase. The increase in $v(T)$ at low temperatures suggests that the increase in $E(T)$ has a dominant effect compared to the increase in $\rho(T)$.

To examine the mechanism underlying the increase in the quality factor, we considered the value used as a figure of merit for mechanical resonators, $f_r \times Q$. From this value, we can infer the dissipation mechanism of the SAW resonator[15]. Figure 5 shows the temperature dependence of $f_r \times Q$. The fitting result shows that $f_r \times Q$ was proportional to $T^{-0.6}$. In the range of the parameter $R_{\text{line}}$ (0.5 to 1.5 Ω) used in this analysis, the power of $T$ should be in the range of -0.5 to -0.9. The temperature dependence of $f_r \times Q$ suggests that the contributions from the impurity scattering and boundary scattering, which are less sensitive to temperature variations, were small[16], and that temperature-dependent phonon-phonon scattering was dominant. AlN and diamond are insulators, and the electron-phonon scattering was considered negligible here.

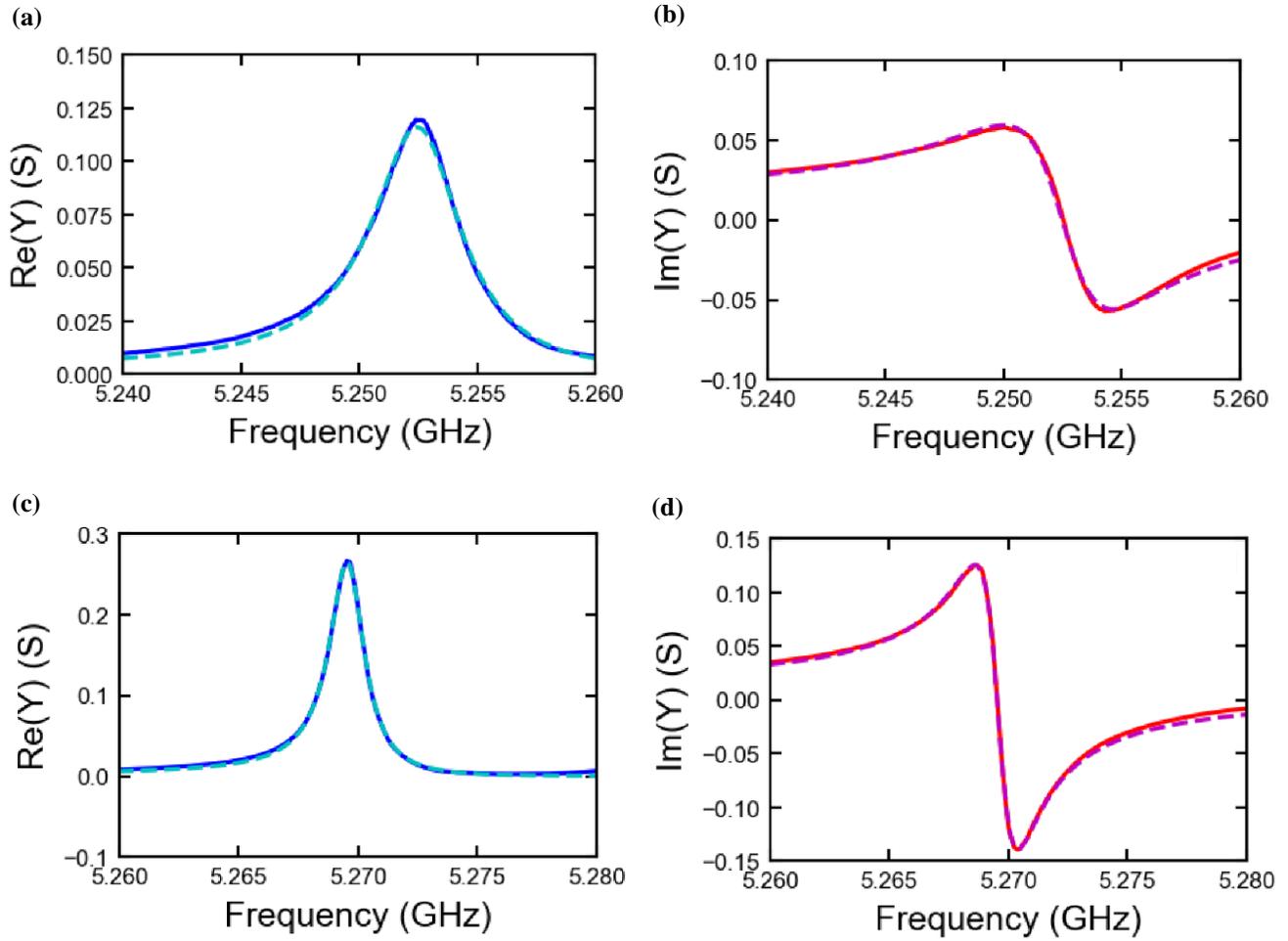

Fig. 3. (a) Real and (b) imaginary parts of admittance at room temperature (300 K) and (c) real and (d) imaginary parts of admittance at 5 K. Solid lines are experimental data and dashed lines are fitting curves.

TABLE I. Parameters obtained from fitting of admittance at various temperatures are shown. Due to the temperature correction procedure we used, $R_{line}$= 1 Ω, $L_{line}$= 2.25 nH, and $C_0$=1.66 fF were treated as constants.

| T (K) | $R_0$ (Ω) | $R_m$ (Ω) | $L_m$ (nH) | $C_m$ (fF) | $f_r$ (GHz) |
|---|---|---|---|---|---|
| 300 | 14.4 | 14.8 | 831 | 1.10 | 5.2648 |
| 250 | 14.1 | 13.0 | 864 | 1.06 | 5.2691 |
| 200 | 13.7 | 11.1 | 879 | 1.04 | 5.2730 |
| 150 | 13.8 | 8.97 | 903 | 1.01 | 5.2763 |
| 100 | 13.7 | 6.73 | 919 | 0.990 | 5.2775 |
| 75 | 13.5 | 5.67 | 927 | 0.981 | 5.2789 |
| 50 | 13.3 | 5.03 | 932 | 0.975 | 5.2798 |
| 25 | 13.2 | 3.72 | 941 | 0.965 | 5.2798 |
| 5 | 13.5 | 1.51 | 944 | 0.962 | 5.2801 |

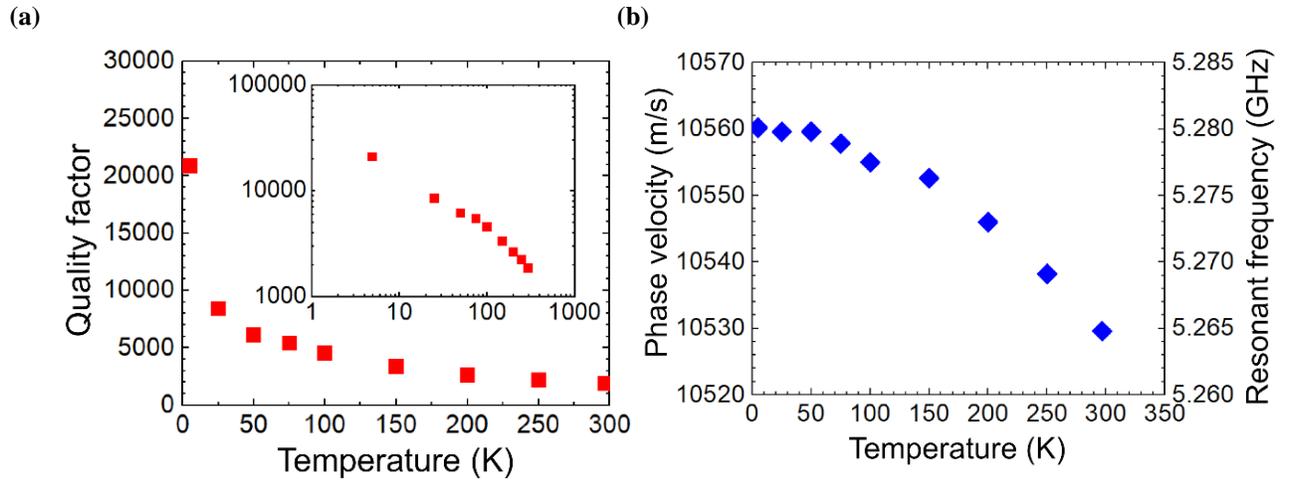

Fig. 4 (a) Temperature dependence of the SAW resonator's quality factor. Inset plots are shown in both logarithmic scales. (b) Temperature dependence of the phase velocity (resonant frequency) of the SAW.

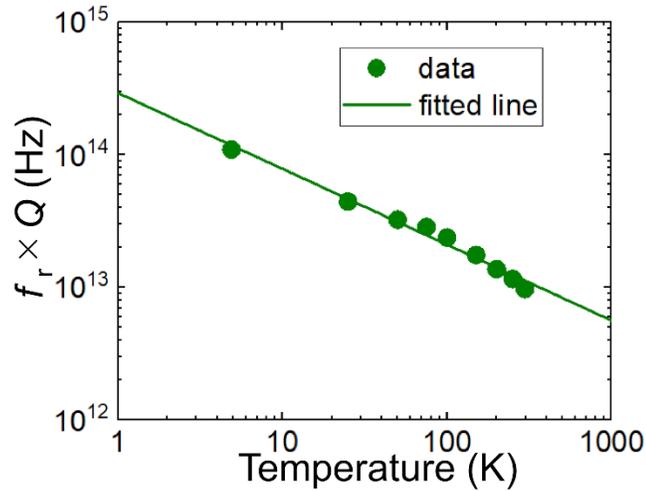

Fig. 5. Temperature variation of the product of resonant frequency and quality factor. The slope of the fitted line is $T^{-0.6}$.

In the case of phonon-phonon scattering, the scattering mechanism can be classified into two categories based on the product of the phonon relaxation time $\tau$ and the SAW angular frequency $\omega$ (= $2\pi f_r$). $f_r \times Q$ are expressed differently in these two regimes as

$$f_r \times Q = \frac{\rho V_a [1+(\omega\tau)^2]}{2\pi C_v \gamma^2 \tau T} \quad (\omega\tau < 1), \tag{4a}$$

$$f_r \times Q = \frac{15\rho V_a^5 h^3 f_r}{\pi^5 k_B^4 \gamma^2 T^4} \quad (\omega\tau > 1), \tag{4b}$$

where $V_a$ is the speed of sound, $C_v$ is the specific heat at constant volume, $\gamma$ is the Grüneisen parameter, $h$ is the Planck constant, and $k_B$ is the Boltzmann constant. When the wavelength of the SAW is longer than the phonon mean free path ($\omega\tau < 1$), it is called the Akheiser region, where $f_r \times Q$ changes according to $f_r \cdot Q \propto 1/(C_v \tau \gamma^2 T)$. When the wavelength of the SAW is shorter than the phonon mean free path ($\omega\tau > 1$), it is called the Landau-Rumer region, and the $f_r \times Q$ changes according to $f_r \times Q \propto 1/(\gamma^2 T^4)$ as in Equation (4b). The measured temperature range was below the Debye temperature of AlN and diamond[17,18], so we assumed $C_v \propto T^3$. We also assumed $\tau \propto T^{-3.4}$ [19]. Assuming that $\gamma$ is a constant that does not change with temperature[20], the temperature dependence can be written as $f_r \times Q \propto T^{-0.6}$ in the Akheiser region and $f_r \times Q \propto T^{-4}$ in the Landau-Rumer region. The obtained $f_r \times Q$ varied with temperature according to $T^{-0.6}$, suggesting that AlN/diamond is in the Akheiser region at 5 GHz. Strictly speaking, $\gamma$ also varies with temperature[21,22], which needs to be taken into account, but the temperature variation in $f_r \times Q$ can be roughly explained by $T^{-0.6}$, so the effect of temperature variation in $\gamma$ was considered small.

At a frequency of 5 GHz, diamond in bulk samples falls within the Landau-Rumer region[15,19]. The present sample could be in the Akheiser region because the phonon relaxation in the AlN thin film deposited on diamond is dominant; in the Akheiser region, phonon lifetimes are short, and it is impossible to observe single phonon-matter interactions. Hence, to prolong phonon lifetime and explore quantum phonon-matter interactions in AlN/diamond SAW devices, future experiments can be conducted at lower temperatures utilizing dilution refrigerators. Additionally, the use of epitaxially grown single-crystal AlN and the minimization of the volume of AlN in the resonator could also be considered.

## IV. CONCLUSION

Reflectance spectra were measured in the range of 5-300 K to investigate the temperature characteristics of the quality factor and $f_r$ of the AlN/diamond heterostructure SAW resonator. The quality factor increased with cooling, mainly due to the suppression of the phonon-phonon scattering with decreasing temperature. The temperature dependence of $f_r \times Q$ suggested that the AlN/diamond SAW resonator is in the Akheiser region at 5 GHz and that the quality factor can be improved by further cooling. In addition, the crystalline structure of AlN and the reduction of the volume fraction in the resonator may be effective in extending the phonon lifetime in the SAW resonator, which is an important guideline for the design of future quantum SAW devices.

## ACKNOWLEDGMENTS

This work is supported by a Japan Science and Technology Agency (JST) Moonshot R&D grant (JPMJMS2062) and by a JST CREST grant (JPMJCR1773). We also acknowledge the Ministry of Internal Affairs and Communications (MIC) for funding, research and development for construction of a global quantum cryptography network (JPMI00316) and the Japan Society for the Promotion of Science (JSPS) Grants-in-Aid for Scientific Research (20H05661, 20K20441).

## AUTHOR DECLARATIONS

**Conflict of Interest**

The authors have no conflicts to disclose.